\pdfoutput=1
\documentclass[12pt,a4paper]{article}
\usepackage[a4paper,margin=2.6cm]{geometry}
\usepackage[T1]{fontenc}
\usepackage[utf8]{inputenc}
\usepackage{mathptmx}              
\usepackage{textcomp}
\DeclareUnicodeCharacter{2212}{\ensuremath{-}}
\DeclareUnicodeCharacter{2248}{\ensuremath{\approx}}
\DeclareUnicodeCharacter{03B2}{\ensuremath{\beta}}
\DeclareUnicodeCharacter{03B5}{\ensuremath{\varepsilon}}
\DeclareUnicodeCharacter{03BC}{\textmu}
\DeclareUnicodeCharacter{03C0}{\ensuremath{\pi}}
\DeclareUnicodeCharacter{03C6}{\ensuremath{\varphi}}
\usepackage[final,protrusion=false,expansion=false]{microtype}
\DisableLigatures{encoding = *, family = *}   
\usepackage{graphicx}
\usepackage{booktabs}
\usepackage{xcolor}
\definecolor{linkblue}{HTML}{0563C1}
\usepackage{titlesec}
\usepackage[hidelinks,pdfencoding=auto,bookmarksnumbered=false,bookmarksopen=true]{hyperref}
\hypersetup{pdftitle={DIAL-ST proposal: Dedicated International Accelerator Laboratory for Space Travel-related radiation research},
  pdfauthor={Dilara Akturk, Arda Goroglu, Saleh Sultansoy, Zahida Sultanova},
  pdfkeywords={Cosmic Radiation, Accelerators, Galactic Cosmic Rays, Solar Particle Events, Biological Pathways}}
\titleformat{\section}{\fontsize{13}{16}\selectfont\bfseries}{}{0pt}{}
\titlespacing*{\section}{0pt}{20pt}{0pt}
\makeatletter\@secpenalty=0\makeatother
\begin{document}
{\centering{\fontsize{16}{20}\selectfont\bfseries DIAL-ST proposal: Dedicated International Accelerator\\Laboratory for Space Travel-related radiation research\par}
\vspace{20pt}
Dilara Akturk\textsuperscript{1}, Arda Goroglu\textsuperscript{1*}, Saleh Sultansoy\textsuperscript{1}†, Zahida Sultanova\textsuperscript{2}\par
\textsuperscript{1}TOBB University of Economics and Technology, Ankara, Türkiye\par
\textsuperscript{2}School of Biological Sciences, University of East Anglia, Norwich, UK\par
*Corresponding Author: \href{mailto:agoroglu@etu.edu.tr}{\textcolor{linkblue}{\underline{agoroglu@etu.edu.tr}}}\par
†Member of ATLAS, LHeC \& FCC collaborations at CERN\par
\par
}
\vspace{14pt}{\centering\fontsize{13}{16}\selectfont\bfseries Abstract\par}\nopagebreak
\pdfbookmark[1]{Abstract}{abstract}
The 21\textsuperscript{st} century is expected to mark the era of human colonization of the Solar System, beginning with permanent settlements on the Moon and subsequently on Mars. Within this context, understanding the biological and technological effects of cosmic radiation has become a critical challenge for the long-term sustainability of space habitats. Current accelerator-based studies have significantly contributed to space radiobiology; however, existing facilities remain limited in their ability to simulate the space radiation environment. To overcome these scientific challenges and current experimental limitations, we propose the \textbf{D}edicated \textbf{I}nternational \textbf{A}ccelerator \textbf{L}aboratory for \textbf{S}pace \textbf{T}ravel-related radiation research (DIAL-ST) as a multi-branch accelerator complex capable of generating simultaneous and tuneable particle beams that can replicate galactic cosmic rays and solar particle events under controlled laboratory conditions. The aim of DIAL-ST is to provide an accurate simulation of mixed-field cosmic radiation, while also contributing to multidisciplinary research in space environments across biology, medicine, physics and technological sciences.\par
\vspace{34pt}{\fontsize{13}{16}\selectfont\bfseries Contents:\par}\vspace{6pt}
1. Introduction\par\vspace{6pt}
2. Cosmic Rays\par\vspace{6pt}
3. Particle Accelerators as Cosmic Radiation Simulators\par\vspace{6pt}
4. Dedicated International Accelerator Laboratory for Space Travel (DIAL-ST) Proposal\par\vspace{6pt}
5. Required Beam Intensities\par\vspace{6pt}
6. Conclusion\par\vspace{6pt}
\clearpage
\section*{1. Introduction}\addcontentsline{toc}{section}{1. Introduction}
The 21\textsuperscript{st} century is anticipated to mark the era of human colonization of the Solar System. In the initial phase, the establishment of permanent settlements on the Moon, followed by colonies on Mars, is envisioned [1]. Within this context, investigating the effects of cosmic radiation on both biological organisms and technological systems holds exceptional importance, as radiation exposure represents one of the most critical challenges to the long-term sustainability of space habitats [2].\par
Space radiation presents a multifaceted challenge for crew health and mission success, particularly given the methodological disparity between how ground-based radiobiology experiments are conducted and the complex, multi-directional exposure patterns characteristic of deep-space environments. Recent experimental and theoretical studies indicate that conventional laboratory studies often rely on monoenergetic beams, sequential irradiation protocols or high-dose-rate exposures that are not representative of the low-dose-rate, mixed-field, and time-variable spectra encountered in the space radiation environment [2–5].\par
A central focus of current research is the need to realistically represent galactic cosmic radiation (GCR) and solar particle events (SPEs), which are classified as harmful cosmic radiation, under controlled conditions in laboratory simulations. GCR consists of protons (≈ 87–90\%), helium nuclei (alpha particles, ≈ 10–12\%) and heavier ions (≈ 1\%), together with electrons and positrons (≈ 1–2\%) [4, 6–8], whereas SPEs consist largely of protons with energies of up to about 1 GeV [7, 9].\par
Thus far, accelerator-based research has mainly focused on understanding how cosmic radiation impacts living organisms rather than on finding ways to help them resist the radiation. In a recent paper [10], we focused on the biological effects of cosmic radiation and proposed possible protective measures through biological interventions. We proposed that biological response pathways responsible for coping with stressors on Earth can provide effective solutions for protection against the stress caused by cosmic radiation. To study the biological effects of cosmic radiation and explore potential ways to protect against it, establishing the Dedicated International Accelerator Laboratory for Space Travel-related radiation research (DIAL-ST), which can effectively simulate cosmic radiation, will be very useful [10].\par
In this study, we introduce a conceptual blueprint for a four-branch, modular accelerator-based laboratory architecture intended to simulate space radiation with higher accuracy and experimental realism. In section 2, we give brief information about cosmic rays. Information about the current particle accelerators used for cosmic radiation simulations is given in section 3. The methodology and conceptual design of the proposed accelerator laboratory are presented in section 4. The beam intensities required to reproduce the cosmic-ray fluxes are estimated in section 5. Finally, section 6 includes our conclusions and recommendations.\par
\section*{2. Cosmic Rays}\addcontentsline{toc}{section}{2. Cosmic Rays}
The energy distribution of different types of cosmic rays is presented in Figure 1 [6]. Cosmic radiation is composed mainly of protons (≈ 87–90\%), helium nuclei (≈ 10–12\%), heavy ions (≈ 1\%) and electrons and positrons (≈ 1–2\%) [4, 6–8]. Figure 1 shows the spectra as E\textsuperscript{2}dN/dE versus the energy per particle; in this representation their maxima lie between about 2 and 100 GeV per particle, i.e. at about 4–6 GeV per nucleon for protons, helium, carbon and oxygen. The differential particle flux dN/dE itself peaks at lower energies, at about 0.3–0.8 GeV for protons depending on solar activity (force-field parametrization [11]), and the region from about 100 MeV per nucleon to 10–50 GeV per nucleon is of primary importance for space applications.\par
\par\vspace{2pt}\noindent\begin{minipage}{\textwidth}\centering
\includegraphics[width=11.27cm]{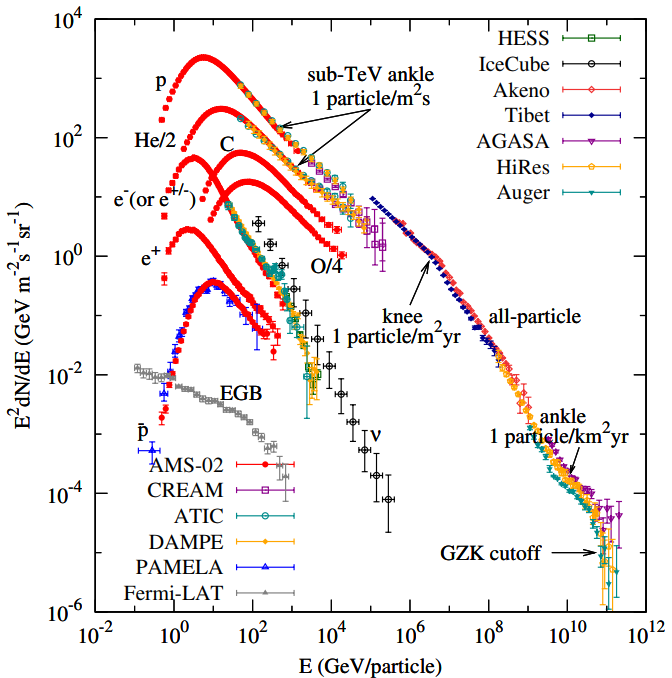}\par\vspace{8pt}
Figure 1. Cosmic-ray energy spectra, shown as E\textsuperscript{2}dN/dE versus the energy per particle E; the helium and oxygen spectra are divided by 2 and 4, respectively. Reproduced from [6] under the CC BY 3.0 licence.\par
\end{minipage}\par\vspace{6pt}
Although the percentage of heavy ions with high energies is relatively low, their biological impact is expected to be more detrimental, as the energy they deposit per unit path length (linear energy transfer, LET) is proportional to the square of their charge at a given velocity [4, 12]. Moreover, above about 10 GeV the flux of cosmic rays decreases with energy as a power law, dN/dE \ensuremath{\propto} E\textsuperscript{−2.7}, up to the knee at a few PeV [6, 12]. At lower energies the spectra turn over because of energy losses and solar modulation, and the ion fluxes at solar maximum can be 3–4 times lower than at solar minimum [8, 12]. During space missions, living organisms are continuously exposed to such radiation sources. Missions conducted aboard the International Space Station (ISS) result in radiation doses significantly higher and more hazardous than those encountered on Earth, while future deep-space missions are predicted to involve even higher doses, with longer exposures [13].\par
\section*{3. Particle Accelerators as Cosmic Radiation Simulators}\addcontentsline{toc}{section}{3. Particle Accelerators as Cosmic Radiation Simulators}
Three major international facilities have so far mainly been used for researching the biological effects of cosmic radiation. The best known among them is the NASA Space Radiation Laboratory (NSRL), which contains a Galactic Cosmic Ray (GCR) simulator capable of delivering 33 sequential ion beams of seven species (H, He, C, O, Si, Ti and Fe) with energies ranging from 20 MeV/u to 1 GeV/u; the facility itself provides protons from 50 MeV to 2.5 GeV and heavier ions from 50 MeV/u up to 1.5 GeV/u (helium, carbon, oxygen) or 1 GeV/u (silicon, titanium, iron) [4, 14–16]. The second major facility, the GSI Helmholtz Centre for Heavy Ion Research (GSI), provides a broad spectrum of ion species ranging from hydrogen to uranium, with energies from about 50 MeV/u up to 1 GeV/u for uranium and 2 GeV/u for light ions such as neon, and up to 4.5 GeV for protons [17, 18]. GSI has also implemented a hybrid active–passive GCR simulator, in which a \textsuperscript{56}Fe beam at 0.35, 0.7 and 1 GeV/u interacts with passive modulators to produce a mixed radiation field, including neutrons, that reproduces the GCR spectrum behind 10 g/cm\textsuperscript{2} of aluminium shielding [8]. The third facility, the Heavy Ion Medical Accelerator in Chiba (HIMAC), delivers ion beams such as helium, carbon, silicon, argon and iron, with energies from 100 MeV/u up to 230 MeV/u for helium, 430 MeV/u for carbon, 800 MeV/u for silicon, 650 MeV/u for argon and 500 MeV/u for iron [19–21]. These facilities have been used to examine the effects of space radiation on various biological characteristics, including cellular function, chromosomal damage, cancer progression, cardiac function, cognitive performance and behaviour [22–27].\par
New capabilities are also emerging. In Europe, the EU-funded HEARTS project provides very-high-energy (above 100 MeV/u) heavy-ion beams for radiation testing of electronics, shielding and radiobiology research at CERN and GSI, and a dedicated facility for electronics testing with ions of up to 100 MeV/u, based on the LEIR synchrotron, has been proposed [28]. At FAIR, beam commissioning is scheduled to start in the second half of 2027, and the SIS100 synchrotron (100 T·m), designed to deliver heavy ions at about 10 GeV/u, is expected to begin operation in late 2028 [29]. In China, the High Intensity heavy-ion Accelerator Facility (HIAF) completed construction and began trial operation in July 2026 [30]. At JINR, the ARIADNA collaboration is commissioning the SIMBO (400–1100 MeV/u, space radiobiology) and ISCRA (150–500 MeV/u, electronics testing) stations at the NICA complex; their diagnostics were first tested with xenon beams in March 2026 [31, 32].\par
Despite the contributions of accelerator-based studies to understanding the effects of cosmic radiation on biological organisms and to protective strategies, experimental set-ups still differ from the true space radiation environment. Solar particle events (SPEs) vary in both intensity and duration, lasting from several hours to days, while galactic cosmic rays are continuously present in the background, with flux variations driven by solar activity [33]. Replicating these conditions is challenging and requires detailed information on the mission profile, including its trajectory and duration. For instance, a mission to Mars would last several hundred days (about 650–920 days in NASA's reference scenarios), whereas a lunar mission lasts from several days for Apollo-type flights (9 days for Artemis II in April 2026) to 30–90 days for the first Gateway and lunar-surface missions [4, 9, 34–36].\par
Accelerator-based acute exposure experiments typically last up to a few hours, whereas chronic exposure studies can be extended over several weeks [37]. Ideally, laboratory simulations should resemble the environment of space, where both the flux and duration of cosmic ray exposure change dynamically over time [33]. Achieving these conditions requires the establishment of dedicated new accelerator laboratories for space radiobiology research or the adaptation of existing accelerators for specialized use in this field [9, 38].\par
\section*{4. Dedicated International Accelerator Laboratory for Space Travel (DIAL-ST) Proposal}\addcontentsline{toc}{section}{4. Dedicated International Accelerator Laboratory for Space Travel (DIAL-ST) Proposal}
A schematic view of the facility for the first stage of DIAL-ST is presented in Figure 2.\par
\par\vspace{2pt}\noindent\begin{minipage}{\textwidth}\centering
\includegraphics[width=14cm]{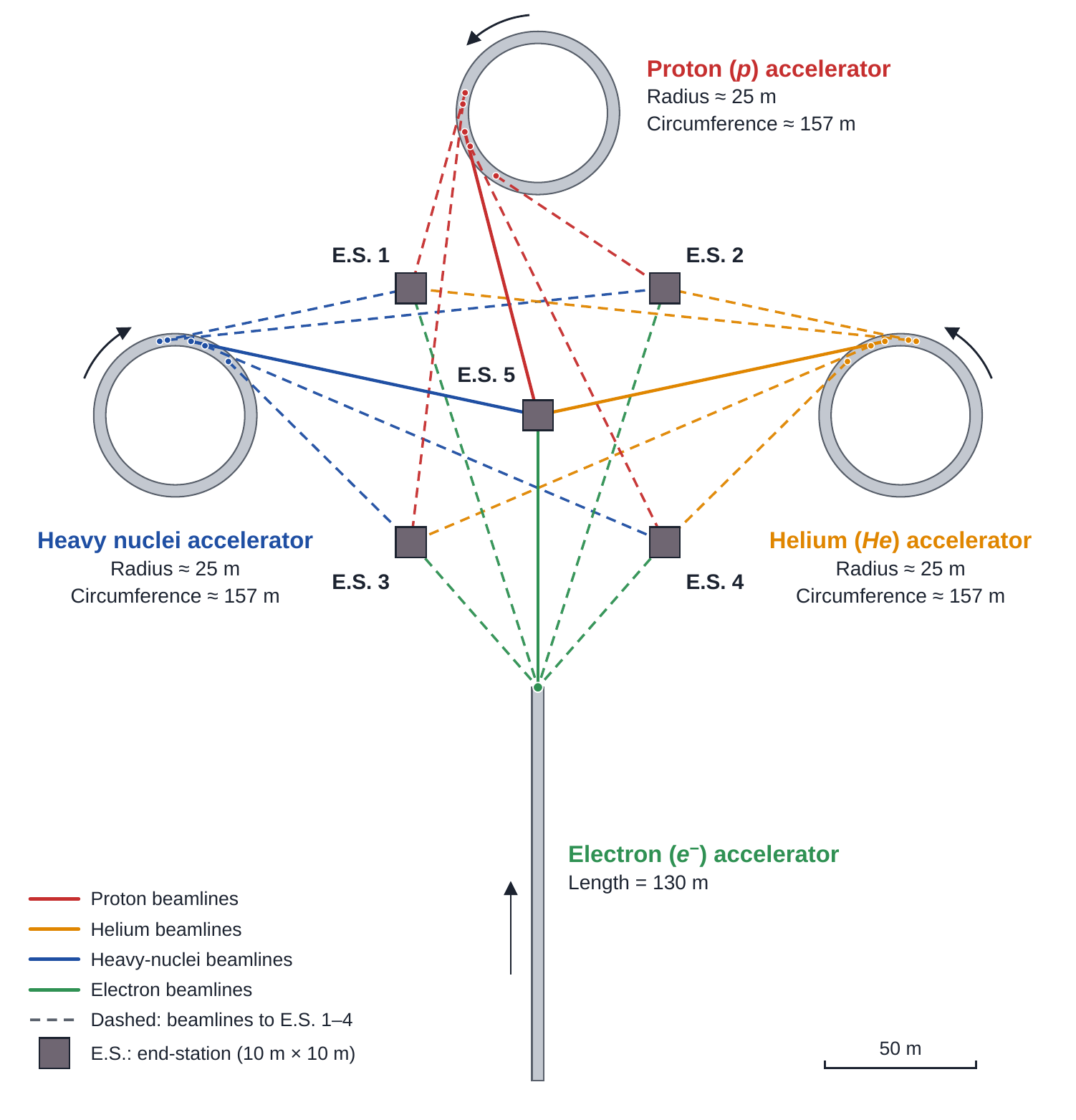}\par\vspace{8pt}
Figure 2. Schematic view of the facility for the first stage of DIAL-ST\par
\end{minipage}\par\vspace{6pt}
The proposed DIAL-ST facility (Figure 2) is designed as a multi-branch accelerator complex, capable of generating simultaneous and tuneable particle fields for space-radiation simulation studies. The facility comprises three circular hadron accelerators and an electron linac, collectively forming a four-branch configuration that enables simultaneous multi-species beam generation.\par
The three independent circular accelerators are modelled with a radius of approximately 25 m (a circumference of 157 m) and the linear accelerator with a length of 130 m. Each of them is dedicated to accelerating a different particle species: protons (\textit{p}), helium ions (\textit{He}), \textit{heavy nuclei} and electrons (\textit{e\textsuperscript{-}}). These circular and linear accelerators are symmetrically arranged around multiple experimental end-stations (E.S. 1–E.S. 5), each with approximate dimensions of 10 m × 10 m. Every particle species (electrons, protons, helium ions and heavy nuclei) can be delivered to each experimental end-station either sequentially or simultaneously depending on the experimental requirements. In future developments, all experimental end-stations can be equipped with a three-dimensional rotational configuration, allowing uniform dose distribution over all surfaces of biological or material specimens during irradiation. Furthermore, each end-station can be architecturally structured as a multi-level building, enabling improved efficiency within the experimental zone.\par
With each branch operating in a different irradiation mode, the system ensures a uniform spatial distribution of the radiation field, precise adjustment of the beam energy, flux and incidence angle at each end-station, and the generation of mixed-field and low-dose-rate beam conditions to simulate space-like exposure characteristics. For the first operational phase it is proposed to begin with beam energies around 1 GeV (1 GeV per nucleon for helium and heavier nuclei), providing a compact and cost-effective prototype configuration. In future studies, the system can be scaled up with additional, larger synchrotrons to reach beam energies of approximately 100 GeV, thereby extending its applicability to a broader range of radiation environments. Additionally, a circular accelerator dedicated to electrons may be considered to reach energies of up to several GeV.\par
The experimental end-station design allows the system to host multiple concurrent experiments, where each station can be equipped with independent dosimetry, biological sample holders and detector arrays. In principle, the number of end-stations can be increased by adopting a multi-tier experimental layout, enabling parallel experiments and research under different radiation field parameters.\par
To extend the operational range of the facility, several upgrade options are foreseen. Lengthening the electron linac section and/or increasing the accelerating gradient (Compact Linear Collider (CLIC) technology or plasma wakefield acceleration (PWFA)) can be used to achieve higher energies; for hadrons, increasing the magnetic field of the 157 m rings gives only a limited gain (protons reach about 3 GeV with 1.6 T dipoles and of the order of 10 GeV with superconducting dipoles of 4–6 T, assuming that the dipoles fill about one third of the circumference), whereas 100 GeV protons require a magnetic rigidity of about 337 T·m and hence additional synchrotrons with circumferences of about 1 km (6 T dipoles) to 4 km (1.6 T dipoles). Such flexibility allows DIAL-ST to simulate a broad range of space radiation spectra, from galactic cosmic rays to solar particle events, within a controlled and reproducible laboratory environment.\par
In terms of cost, for the dimensions specified above and a beam energy of 1 GeV, the estimated cost is approximately \$200 million for each circular accelerator and \$100 million for the linear accelerator, including the general facility infrastructure. The cost of each end-station may vary depending on its design and functionality, ranging from \$20 million to \$30 million. Additionally, the facility includes a total of 20 beamlines (4 beam sources × 5 end-stations), each estimated at approximately \$1–2 million. In total, the estimated cost of the project for the first-stage design, including the beamline systems, is approximately \$0.8–0.9 billion.\par
\section*{5. Required Beam Intensities}\addcontentsline{toc}{section}{5. Required Beam Intensities}
To estimate the beam intensities needed at the end-stations in the first stage, the cosmic-ray intensities measured by AMS-02 [39–42], as compiled in the Cosmic-Ray Database [43], were integrated over the kinetic-energy band 0.5–1.5 GeV (per nucleon for nuclei) around the first-stage energy of 1 GeV. The AMS-02 spectra of nuclei are given as functions of the magnetic rigidity \textit{R} = \textit{pc}/(\textit{Ze}), i.e. the momentum per unit charge, which determines the curvature of the trajectory of a particle in a magnetic field; the limits of the band were therefore converted from kinetic energy to rigidity. For an isotropic radiation field with band-integrated intensity \textit{I}, the omnidirectional flux to be reproduced by the beams is \textit{φ} = 4π\textit{I}, and the particle rate on an irradiated area \textit{S} is \textit{\.{N}} = \textit{φS}. Table 1 lists \textit{I}, \textit{φ} and \textit{\.{N}} for \textit{S} = 1 m\textsuperscript{2} at the natural rate, i.e. without time compression and without beam losses between the accelerator and the sample. In this band the nuclei consist of 87\% protons, 12\% helium and about 1\% heavier nuclei, while electrons and positrons account for 2.6\% of all particles.\par
\par\vspace{4pt}\noindent\begin{minipage}{\textwidth}
Table 1. Cosmic-ray intensity \textit{I} integrated over the kinetic-energy band 0.5–1.5 GeV (per nucleon for nuclei), omnidirectional flux \textit{φ} = 4π\textit{I} and particle rate \textit{\.{N}} on an area of 1 m\textsuperscript{2}. \textit{E} is the corresponding kinetic energy per particle, as on the axis of Figure 1, and \textit{R} the rigidity.\par\vspace{7pt}
{\fontsize{11}{15}\selectfont\begin{tabular*}{\textwidth}{@{}l@{\extracolsep{\fill}}rrrrr@{}}\toprule
\textbf{Species} & \textbf{\textit{E} (GeV)} & \textbf{\textit{R} (GV)} & \textbf{\textit{I} (m\textsuperscript{−2} s\textsuperscript{−1} sr\textsuperscript{−1})} & \textbf{\textit{φ} (cm\textsuperscript{−2} s\textsuperscript{−1})} & \textbf{\textit{\.{N}} (s\textsuperscript{−1})} \\ \midrule
Protons & 0.5–1.5 & 1.09–2.25 & 672 & 0.844 & 8.4 × 10\textsuperscript{3} \\
Helium & 2–6 & 2.17–4.49 & 91.3 & 0.115 & 1.15 × 10\textsuperscript{3} \\
Carbon & 6–18 & 2.17–4.49 & 2.45 & 3.1 × 10\textsuperscript{−3} & 31 \\
Oxygen & 8–24 & 2.17–4.49 & 2.32 & 2.9 × 10\textsuperscript{−3} & 29 \\
Iron & 28–84 & 2.34–4.84 & 0.16 & 2.0 × 10\textsuperscript{−4} & 2.0 \\
Li–Fe, sum of 14 elements &  &  & 8.5 & 1.07 × 10\textsuperscript{−2} & 107 \\
Electrons & 0.5–1.5 &  & 18.7 & 2.35 × 10\textsuperscript{−2} & 235 \\
Positrons & 0.5–1.5 &  & 1.65 & 2.1 × 10\textsuperscript{−3} & 21 \\
\bottomrule\end{tabular*}}
\end{minipage}\par\vspace{8pt}
Since the energy deposition of high-energy positrons in matter is practically the same as that of electrons, the lepton component is represented in this study by the electron beam alone, with its flux scaled to the sum of the electron and positron fluxes.\par
In the first stage each hadron ring is operated in a storage mode: the beam is accelerated to the working energy, kept circulating and extracted slowly and continuously towards the end-stations. When a given fraction of the stored particles has been used, the ring is refilled. For a circumference of 157 m the revolution frequency at 1 GeV per nucleon is \textit{f}\textsubscript{rev} = \textit{βc}/\textit{C} ≈ 1.67 MHz, so that \textit{N} stored particles of charge \textit{Ze} correspond to a circulating current \textit{I}\textsubscript{c} = \textit{ZeNf}\textsubscript{rev}. If a fraction \textit{f} of the stored beam is used between two fills, the time between fills is \textit{T} = \textit{fNε}/(\textit{n\.{N}}), where \textit{n} is the number of end-stations served simultaneously and \textit{ε} the efficiency of extraction and delivery to the irradiated area.\par
Table 2 gives a set of parameters for \textit{n} = 5, \textit{f} = 0.1 and an assumed efficiency \textit{ε} = 0.1 at the natural rate. The stored intensities are chosen such that the circulating currents remain well above the resolution of DC current transformers, about 1 μA [44]. With 10\textsuperscript{10} protons and 10\textsuperscript{9} helium nuclei the rings are refilled about every 40 and 30 minutes, respectively; a refill takes one acceleration cycle, i.e. a few seconds [45]. The heavy-nuclei ring stores one species at a time, so that the species are delivered in turn; with 10\textsuperscript{8} ions per fill each species can supply its share for a rotation period of one hour.\par
\par\vspace{4pt}\noindent\begin{minipage}{\textwidth}
Table 2. Parameters of the hadron rings in the storage mode for five end-stations, \textit{f} = 0.1 and an assumed efficiency \textit{ε} = 0.1 at the natural rate: required rate \textit{n\.{N}}/\textit{ε}, number of stored particles \textit{N}, circulating current \textit{I}\textsubscript{c} and time \textit{T} between fills.\par\vspace{7pt}
{\fontsize{11}{15}\selectfont\begin{tabular*}{\textwidth}{@{}l@{\extracolsep{\fill}}rrrr@{}}\toprule
\textbf{Ring} & \textbf{\textit{n\.{N}}/\textit{ε} (s\textsuperscript{−1})} & \textbf{\textit{N}} & \textbf{\textit{I}\textsubscript{c} (mA)} & \textbf{\textit{T}} \\ \midrule
Protons & 4.2 × 10\textsuperscript{5} & 10\textsuperscript{10} & 2.7 & 40 min \\
Helium & 5.7 × 10\textsuperscript{4} & 10\textsuperscript{9} & 0.54 & 29 min \\
Heavy nuclei (carbon) & 1.5 × 10\textsuperscript{3} & 10\textsuperscript{8} & 0.16 & species in turn \\
Heavy nuclei (iron) & 1.0 × 10\textsuperscript{2} & 10\textsuperscript{8} & 0.70 & species in turn \\
\bottomrule\end{tabular*}}
\end{minipage}\par\vspace{8pt}
The exposures can also be accelerated. We define the time-compression factor \textit{k} as the ratio of the particle rate delivered to the sample to the natural rate given in Table 1: \textit{k} = 1 reproduces the exposure in real time, whereas for \textit{k} \textgreater{} 1 the fluence of a mission of duration \textit{t} is delivered in the time \textit{t}/\textit{k}. For example, with \textit{k} = 100 the fluence of a 920-day Mars mission is delivered in about 9 days. Since the required rates increase in proportion to \textit{k}, the stored intensities have to be increased by the same factor if the time between fills is to remain unchanged; 10\textsuperscript{12} stored protons then correspond to \textit{k} ≈ 100.\par
\section*{6. Conclusion}\addcontentsline{toc}{section}{6. Conclusion}
In this study, we have proposed the development of the Dedicated International Accelerator Laboratory for Space Travel-related radiation research (DIAL-ST) as a next-generation, multi-branch accelerator complex aimed at reproducing the complex radiation conditions encountered in outer space. The DIAL-ST concept integrates circular and linear accelerators to generate simultaneous and tuneable mixed-field particle beams, enabling systematic research into biological, medical, physical and technological responses to galactic cosmic rays and solar particle events under controlled laboratory conditions.\par
The proposed facility is designed to overcome the limitations of existing accelerator-based radiation platforms, which have a limited capability to reproduce the multi-directional, low-dose-rate and time-variable spectra characteristic of the space radiation environment. By offering the capability to perform simultaneous irradiation with multiple particle species, including protons (\textit{p}), helium ions (\textit{He}), \textit{heavy nuclei} and electrons (\textit{e\textsuperscript{-}}), the DIAL-ST system will bridge the experimental gap between conventional ground-based studies and the complex, true space radiation environment in deep-space missions. The interdisciplinary nature of DIAL-ST will also facilitate collaboration among biologists, physicists, medical professionals and engineers to develop comprehensive models for radiation exposure and evaluate radioprotective mechanisms at both biological and technological levels.\par
The total estimated cost of the 1 GeV beam system is approximately \$700 million, and that of the five experimental end-stations and 20 beamlines is approximately \$120–190 million. The result is a scalable first-phase prototype that can be upgraded to higher energy levels and expanded into more complex configurations.\par
In conclusion, the realization of DIAL-ST would mark an important step toward advancing space radiobiology and radiation protection research. Beyond its direct scientific advantage, this facility would serve as a cornerstone for preparing future human and robotic missions for sustained operations in off-Earth environments. Through such international and multidisciplinary collaboration, DIAL-ST could play a significant role in enabling the safe presence of living systems beyond Earth.\par
\section*{References}\addcontentsline{toc}{section}{References}
\begin{list}{}{\setlength{\leftmargin}{2.7em}\setlength{\labelwidth}{2.2em}\setlength{\labelsep}{0.5em}\setlength{\itemsep}{8pt}\setlength{\parsep}{0pt}\setlength{\topsep}{0pt}\setlength{\partopsep}{0pt}\interlinepenalty=10000 \emergencystretch=4em }
\item[{[1]}\hfill] Neukart, F. (2024). Towards sustainable horizons: A comprehensive blueprint for Mars colonization. Heliyon, 10(4), e26180. \href{https://doi.org/10.1016/j.heliyon.2024.e26180}{https://\penalty400 doi\penalty700 .org\allowbreak /10\allowbreak .1016\allowbreak /j\allowbreak .heliyon\allowbreak .2024.e26180}
\item[{[2]}\hfill] Furukawa, S., Nagamatsu, A., Nenoi, M., et al. (2020). Space Radiation Biology for “Living in Space”. BioMed Research International, 2020, 4703286. \href{https://doi.org/10.1155/2020/4703286}{https://\penalty400 doi\penalty700 .org\allowbreak /10\allowbreak .1155\allowbreak /2020/4703286}
\item[{[3]}\hfill] Alaghband, Y., Klein, P. M., Kramár, E. A., et al. (2023). Galactic cosmic radiation exposure causes multifaceted neurocognitive impairments. Cellular and Molecular Life Sciences, 80(1), 29. \href{https://doi.org/10.1007/s00018-022-04666-8}{https://\penalty400 doi\penalty700 .org\allowbreak /10\allowbreak .1007\allowbreak /s00018\allowbreak -022-04666-8}
\item[{[4]}\hfill] Simonsen, L. C., Slaba, T. C., Guida, P., \& Rusek, A. (2020). NASA’s first ground-based Galactic Cosmic Ray Simulator: Enabling a new era in space radiobiology research. PLOS Biology, 18(5), e3000669. \href{https://doi.org/10.1371/journal.pbio.3000669}{https://\penalty400 doi\penalty700 .org\allowbreak /10\allowbreak .1371\allowbreak /journal\allowbreak .pbio.3000669}
\item[{[5]}\hfill] Kimura, A., Matsufuji, N., Hiroki, A., Seito, H., \& Taguchi, M. (2018). Development of high-sensitivity intra-corporeal catheter-type liquid dosimeter for radiotherapy. Biomedical Physics \& Engineering Express, 4(5), 055005. \href{https://doi.org/10.1088/2057-1976/aad395}{https://\penalty400 doi\penalty700 .org\allowbreak /10\allowbreak .1088\allowbreak /2057\allowbreak -1976/aad395}
\item[{[6]}\hfill] Piazzoli, B. D’E., Liu, S.-M., della Volpe, D., et al. (2022). Chapter 4 Cosmic-Ray Physics. Chinese Physics C, 46(3), 030004. \href{https://doi.org/10.1088/1674-1137/ac3faa}{https://\penalty400 doi\penalty700 .org\allowbreak /10\allowbreak .1088\allowbreak /1674\allowbreak -1137/ac3faa}
\item[{[7]}\hfill] Chancellor, J. C., Scott, G. B. I., \& Sutton, J. P. (2014). Space Radiation: The Number One Risk to Astronaut Health beyond Low Earth Orbit. Life, 4(3), 491–510. \href{https://doi.org/10.3390/life4030491}{https://\penalty400 doi\penalty700 .org\allowbreak /10\allowbreak .3390\allowbreak /life4030491}
\item[{[8]}\hfill] Lunati, L., Pierobon, E., Weber, U., Wagner, T., Pfuhl, T., Durante, M., \& Schuy, C. (2026). Hybrid active–passive Galactic Cosmic Ray simulator: In-silico design and optimization. Life Sciences in Space Research, 51, 94–111. \href{https://doi.org/10.1016/j.lssr.2026.02.003}{https://\penalty400 doi\penalty700 .org\allowbreak /10\allowbreak .1016\allowbreak /j\allowbreak .lssr\allowbreak .2026.02.003}
\item[{[9]}\hfill] Fogtman, A., Baatout, S., Baselet, B., Berger, T., Hellweg, C. E., Jiggens, P., La Tessa, C., Narici, L., Nieminen, P., Sabatier, L., Santin, G., Schneider, U., Straube, U., Tabury, K., Tinganelli, W., Walsh, L., \& Durante, M. (2023). Towards sustainable human space exploration—priorities for radiation research to quantify and mitigate radiation risks. npj Microgravity, 9(1), 8. \href{https://doi.org/10.1038/s41526-023-00262-7}{https://\penalty400 doi\penalty700 .org\allowbreak /10\allowbreak .1038\allowbreak /s41526\allowbreak -023-00262-7}
\item[{[10]}\hfill] Sultanova, Z., \& Sultansoy, S. (2024). Beyond Earthly Limits: Protection against Cosmic Radiation through Biological Response Pathways. arXiv preprint arXiv:2405.12151. \href{https://doi.org/10.48550/arXiv.2405.12151}{https://\penalty400 doi\penalty700 .org\allowbreak /10\allowbreak .48550\allowbreak /arXiv\allowbreak .2405.12151}
\item[{[11]}\hfill] Usoskin, I. G., Gil, A., Kovaltsov, G. A., Mishev, A. L., \& Mikhailov, V. V. (2017). Heliospheric modulation of cosmic rays during the neutron monitor era: Calibration using PAMELA data for 2006–2010. Journal of Geophysical Research: Space Physics, 122(4), 3875–3887. \href{https://doi.org/10.1002/2016JA023819}{https://\penalty400 doi\penalty700 .org\allowbreak /10\allowbreak .1002\allowbreak /2016JA023819}
\item[{[12]}\hfill] Takahashi, F., et al. (Particle Data Group). (2026). Review of Particle Physics. International Journal of Modern Physics A, 41, 2630011. \href{https://doi.org/10.1142/S0217751X26300115}{https://\penalty400 doi\penalty700 .org\allowbreak /10\allowbreak .1142\allowbreak /S0217751X26300115}
\item[{[13]}\hfill] Thirsk, R., Kuipers, A., Mukai, C., \& Williams, D. (2009). The space-flight environment: the International Space Station and beyond. Canadian Medical Association Journal, 180(12), 1216–1220. \href{https://doi.org/10.1503/cmaj.081125}{https://\penalty400 doi\penalty700 .org\allowbreak /10\allowbreak .1503\allowbreak /cmaj.081125}
\item[{[14]}\hfill] NSRL. (2026). Ion Species and Energies – NSRL Physics User Guide. Brookhaven National Laboratory. \href{https://wpw.bnl.gov/nsrl/?p=73}{https://\penalty400 wpw\penalty700 .bnl\penalty700 .gov\allowbreak /nsrl/?p=73}
\item[{[15]}\hfill] Huff, J. L., Poignant, F., Rahmanian, S., Khan, N., Blakely, E. A., Britten, R. A., Chang, P., Fornace, A. J., Hada, M., Kronenberg, A., Norman, R. B., Patel, Z. S., Shay, J. W., Weil, M. M., Simonsen, L. C., \& Slaba, T. C. (2023). Galactic cosmic ray simulation at the NASA space radiation laboratory – Progress, challenges and recommendations on mixed-field effects. Life Sciences in Space Research, 36, 90–104. \href{https://doi.org/10.1016/j.lssr.2022.09.001}{https://\penalty400 doi\penalty700 .org\allowbreak /10\allowbreak .1016\allowbreak /j\allowbreak .lssr\allowbreak .2022.09.001}
\item[{[16]}\hfill] Norbury, J. W., Schimmerling, W., Slaba, T. C., Azzam, E. I., Badavi, F. F., Baiocco, G., Benton, E., Bindi, V., Blakely, E. A., Blattnig, S. R., Boothman, D. A., Borak, T. B., Britten, R. A., Curtis, S., Dingfelder, M., Durante, M., Dynan, W. S., Eisch, A. J., Robin Elgart, S., … Zeitlin, C. J. (2016). Galactic cosmic ray simulation at the NASA Space Radiation Laboratory. Life Sciences in Space Research, 8, 38–51. \href{https://doi.org/10.1016/j.lssr.2016.02.001}{https://\penalty400 doi\penalty700 .org\allowbreak /10\allowbreak .1016\allowbreak /j\allowbreak .lssr\allowbreak .2016.02.001}
\item[{[17]}\hfill] Durante, M., Reitz, G., \& Angerer, O. (2010). Space radiation research in Europe: Flight experiments and ground-based studies. Radiation and Environmental Biophysics, 49(3), 295–302. \href{https://doi.org/10.1007/s00411-010-0300-6}{https://\penalty400 doi\penalty700 .org\allowbreak /10\allowbreak .1007\allowbreak /s00411\allowbreak -010-0300-6}
\item[{[18]}\hfill] GSI. (2026). SIS18 Beam Properties. GSI Helmholtzzentrum für Schwerionenforschung. \href{https://www.gsi.de/en/work/accelerator_operations/accelerators/sis18/beam_properties}{https://\penalty400 www\penalty700 .gsi\penalty700 .de\allowbreak /en\allowbreak /work\allowbreak /accelerator\allowbreak \_operations\allowbreak /accelerators\allowbreak /sis18\allowbreak /beam\allowbreak \_properties}
\item[{[19]}\hfill] Sato, K., et al. (1994). Status Report on HIMAC. Proceedings of the 4th European Particle Accelerator Conference (EPAC’94), 412–416. \href{https://proceedings.jacow.org/e94/PDF/EPAC1994_0412.PDF}{https://\penalty400 proceedings\penalty700 .jacow\penalty700 .org\allowbreak /e94\allowbreak /PDF\allowbreak /EPAC1994\allowbreak \_0412.PDF}
\item[{[20]}\hfill] QST. (2025). Call for HIMAC Machine-Time Request and Terms and Conditions of HIMAC Usage (V25-I). National Institutes for Quantum Science and Technology. \href{https://www.qst.go.jp/uploaded/attachment/43198.pdf}{https://\penalty400 www\penalty700 .qst\penalty700 .go\penalty700 .jp\allowbreak /uploaded\allowbreak /attachment\allowbreak /43198.pdf}
\item[{[21]}\hfill] Miller, J. (2019). 5.1 Space Radiation Physics and Biology. Radioisotopes, 68(6), 403–405. \href{https://doi.org/10.3769/radioisotopes.68.403}{https://\penalty400 doi\penalty700 .org\allowbreak /10\allowbreak .3769\allowbreak /radioisotopes.68.403}
\item[{[22]}\hfill] Huang, E. G., Lin, Y., Ebert, M., Ham, D. W., Zhang, C. Y., \& Sachs, R. K. (2019). Synergy theory for murine Harderian gland tumours after irradiation by mixtures of high-energy ionized atomic nuclei. Radiation and Environmental Biophysics, 58(2), 151–166. \href{https://doi.org/10.1007/s00411-018-00774-x}{https://\penalty400 doi\penalty700 .org\allowbreak /10\allowbreak .1007\allowbreak /s00411\allowbreak -018-00774-x}
\item[{[23]}\hfill] Kiffer, F. C., Luitel, K., Tran, F. H., Patel, R. A., Guzman, C. S., Soler, I., Xiao, R., Shay, J. W., Yun, S., \& Eisch, A. J. (2022). Effects of a 33-ion sequential beam galactic cosmic ray analog on male mouse behavior and evaluation of CDDO-EA as a radiation countermeasure. Behavioural Brain Research, 419, 113677. \href{https://doi.org/10.1016/j.bbr.2021.113677}{https://\penalty400 doi\penalty700 .org\allowbreak /10\allowbreak .1016\allowbreak /j\allowbreak .bbr\allowbreak .2021.113677}
\item[{[24]}\hfill] Nemec-Bakk, A. S., Sridharan, V., Desai, P., Landes, R. D., Hart, B., Allen, A. R., \& Boerma, M. (2023). Effects of simulated 5-ion galactic cosmic radiation on function and structure of the mouse heart. Life, 13(3), 795. \href{https://doi.org/10.3390/life13030795}{https://\penalty400 doi\penalty700 .org\allowbreak /10\allowbreak .3390\allowbreak /life13030795}
\item[{[25]}\hfill] Raber, J., Fuentes Anaya, A., Torres, E. R. S., Lee, J., Boutros, S., Grygoryev, D., Hammer, A., Kasschau, K. D., Sharpton, T. J., Turker, M. S., \& Kronenberg, A. (2020). Effects of six sequential charged particle beams on behavioral and cognitive performance in B6D2F1 female and male mice. Frontiers in Physiology, 11, 959. \href{https://doi.org/10.3389/fphys.2020.00959}{https://\penalty400 doi\penalty700 .org\allowbreak /10\allowbreak .3389\allowbreak /fphys\allowbreak .2020.00959}
\item[{[26]}\hfill] Roggan, M. D., Kronenberg, J., Wollert, E., Hoffmann, S., Nisar, H., Konda, B., Diegeler, S., Liemersdorf, C., \& Hellweg, C. E. (2023). Unraveling astrocyte behavior in the space brain: Radiation response of primary astrocytes. Frontiers in Public Health, 11, 1063250. \href{https://doi.org/10.3389/fpubh.2023.1063250}{https://\penalty400 doi\penalty700 .org\allowbreak /10\allowbreak .3389\allowbreak /fpubh\allowbreak .2023.1063250}
\item[{[27]}\hfill] Yoshida, K., Hada, M., Kizu, A., Kitada, K., Eguchi-Kasai, K., Kokubo, T., Teramura, T., Yano, S., Suzuki, H. H., Watanabe, H., Kondoh, G., Nagamatsu, A., Saganti, P., Cucinotta, F. A., \& Morita, T. (2022). Comparison of biological measurement and physical estimates of space radiation in the International Space Station. Heliyon, 8(8), e10266. \href{https://doi.org/10.1016/j.heliyon.2022.e10266}{https://\penalty400 doi\penalty700 .org\allowbreak /10\allowbreak .1016\allowbreak /j\allowbreak .heliyon\allowbreak .2022.e10266}
\item[{[28]}\hfill] Cortés García, E. C., García Alía, R., Fraser, M. A., Słupecki, M., Waets, A., \& Widorski, M. (2026). A European high-capacity 100 MeV-per-nucleon heavy-ion facility for electronics irradiation based at CERN: Conceptual Design Report. arXiv preprint arXiv:2608.26369.
\item[{[29]}\hfill] Reimann, S., Herlert, A., Schröder, C., et al. (2025). FAIR commissioning - towards first science. Proceedings of IPAC2025, Taipei, Taiwan, THBN2. arXiv:2510.14948.
\item[{[30]}\hfill] Chinese Academy of Sciences. (2026, July 21). China's High Intensity Heavy-ion Accelerator Facility Completes Construction, Begins Trial Operations. \href{https://english.cas.cn/newsroom/news-updates/202607/t20260717_1178561.shtml}{https://\penalty400 english\penalty700 .cas\penalty700 .cn\allowbreak /newsroom\allowbreak /news\allowbreak -updates\allowbreak /202607\allowbreak /t20260717\allowbreak \_1178561.shtml}
\item[{[31]}\hfill] Filatov, G., Slivin, A., Agapov, A., et al. (2022). Beam Lines and Stations for Applied Research Based on Ion Beams Extracted from Nuclotron. Proceedings of IPAC'22, 3096–3098. \href{https://doi.org/10.18429/JACoW-IPAC2022-THPOMS054}{https://\penalty400 doi\penalty700 .org\allowbreak /10\allowbreak .18429\allowbreak /JACoW\allowbreak -IPAC2022\allowbreak -THPOMS054}
\item[{[32]}\hfill] JINR. (2026, April 16). Results of Run 1 at NICA Accelerator Complex. Dubna, No. 16 (4814). \href{https://jinrmag.jinr.ru/eng/2026/16/ip16.htm}{https://\penalty400 jinrmag\penalty700 .jinr\penalty700 .ru\allowbreak /eng\allowbreak /2026\allowbreak /16\allowbreak /ip16.htm}
\item[{[33]}\hfill] Durante, M., \& Cucinotta, F. A. (2011). Physical basis of radiation protection in space travel. Reviews of Modern Physics, 83(4), 1245–1281. \href{https://doi.org/10.1103/RevModPhys.83.1245}{https://\penalty400 doi\penalty700 .org\allowbreak /10\allowbreak .1103\allowbreak /RevModPhys.83.1245}
\item[{[34]}\hfill] Horneck, G., Facius, R., Reichert, M., Rettberg, P., Seboldt, W., Manzey, D., Comet, B., Maillet, A., Preiss, H., Schauer, L., Dussap, C. G., Poughon, L., Belyavin, A., Reitz, G., Baumstark-Khan, C., \& Gerzer, R. (2006). HUMEX, a study on the survivability and adaptation of humans to long duration exploratory missions, part II: Missions to Mars. Advances in Space Research, 38(4), 752–759. \href{https://doi.org/10.1016/j.asr.2005.06.072}{https://\penalty400 doi\penalty700 .org\allowbreak /10\allowbreak .1016\allowbreak /j\allowbreak .asr\allowbreak .2005.06.072}
\item[{[35]}\hfill] Märki, A. (2020). Radiation Analysis for Moon and Mars Missions. International Journal of Astrophysics and Space Science, 8(3), 16–26. \href{https://doi.org/10.11648/j.ijass.20200803.11}{https://\penalty400 doi\penalty700 .org\allowbreak /10\allowbreak .11648\allowbreak /j\allowbreak .ijass\allowbreak .20200803.11}
\item[{[36]}\hfill] NASA. (2026). Artemis II. \href{https://www.nasa.gov/artemis-ii}{https://\penalty400 www\penalty700 .nasa\penalty700 .gov\allowbreak /artemis-ii}
\item[{[37]}\hfill] Luitel, K., Siteni, S., Barron, S., \& Shay, J. W. (2024). Simulated galactic cosmic radiation-induced cancer progression in mice. Life Sciences in Space Research, 41, 43–51. \href{https://doi.org/10.1016/j.lssr.2024.01.008}{https://\penalty400 doi\penalty700 .org\allowbreak /10\allowbreak .1016\allowbreak /j\allowbreak .lssr\allowbreak .2024.01.008}
\item[{[38]}\hfill] Durante, M., \& Kronenberg, A. (2005). Ground-based research with heavy ions for space radiation protection. Advances in Space Research, 35(2), 180–184. \href{https://doi.org/10.1016/j.asr.2004.12.034}{https://\penalty400 doi\penalty700 .org\allowbreak /10\allowbreak .1016\allowbreak /j\allowbreak .asr\allowbreak .2004.12.034}
\item[{[39]}\hfill] Aguilar, M., et al. (AMS Collaboration). (2015). Precision Measurement of the Proton Flux in Primary Cosmic Rays from Rigidity 1 GV to 1.8 TV with the Alpha Magnetic Spectrometer on the International Space Station. Physical Review Letters, 114, 171103. \href{https://doi.org/10.1103/PhysRevLett.114.171103}{https://\penalty400 doi\penalty700 .org\allowbreak /10\allowbreak .1103\allowbreak /PhysRevLett\allowbreak .114.171103}
\item[{[40]}\hfill] Aguilar, M., et al. (AMS Collaboration). (2015). Precision Measurement of the Helium Flux in Primary Cosmic Rays of Rigidities 1.9 GV to 3 TV with the Alpha Magnetic Spectrometer on the International Space Station. Physical Review Letters, 115, 211101. \href{https://doi.org/10.1103/PhysRevLett.115.211101}{https://\penalty400 doi\penalty700 .org\allowbreak /10\allowbreak .1103\allowbreak /PhysRevLett\allowbreak .115.211101}
\item[{[41]}\hfill] Aguilar, M., et al. (AMS Collaboration). (2017). Observation of the Identical Rigidity Dependence of He, C, and O Cosmic Rays at High Rigidities by the Alpha Magnetic Spectrometer on the International Space Station. Physical Review Letters, 119, 251101. \href{https://doi.org/10.1103/PhysRevLett.119.251101}{https://\penalty400 doi\penalty700 .org\allowbreak /10\allowbreak .1103\allowbreak /PhysRevLett\allowbreak .119.251101}
\item[{[42]}\hfill] Aguilar, M., et al. (AMS Collaboration). (2014). Electron and Positron Fluxes in Primary Cosmic Rays Measured with the Alpha Magnetic Spectrometer on the International Space Station. Physical Review Letters, 113, 121102. \href{https://doi.org/10.1103/PhysRevLett.113.121102}{https://\penalty400 doi\penalty700 .org\allowbreak /10\allowbreak .1103\allowbreak /PhysRevLett\allowbreak .113.121102}
\item[{[43]}\hfill] Maurin, D., Ahlers, M., Dembinski, H., Haungs, A., Mangeard, P.-S., Melot, F., et al. (2023). A cosmic-ray database update: CRDB v4.1. European Physical Journal C, 83, 971. \href{https://doi.org/10.1140/epjc/s10052-023-12092-8}{https://\penalty400 doi\penalty700 .org\allowbreak /10\allowbreak .1140\allowbreak /epjc\allowbreak /s10052\allowbreak -023-12092-8}
\item[{[44]}\hfill] Odier, P., Ludwig, M., \& Thoulet, S. (2009). The DCCT for the LHC Beam Intensity Measurement. Proceedings of DIPAC09, Basel, Switzerland, MOPD45. \href{https://epaper.kek.jp/d09/papers/mopd45.pdf}{https://\penalty400 epaper\penalty700 .kek\penalty700 .jp\allowbreak /d09\allowbreak /papers\allowbreak /mopd45.pdf}
\item[{[45]}\hfill] Weeks, J. M., Simonsen, L. C., \& Guida, P. (2023). The NASA Space Radiation Laboratory. 17th International Congress for Radiation Research, Montréal, Canada, August 27–30, 2023. \href{https://ntrs.nasa.gov/citations/20230005226}{https://\penalty400 ntrs\penalty700 .nasa\penalty700 .gov\allowbreak /citations\allowbreak /20230005226}
\end{list}
\end{document}